 \documentstyle[manuscript,aps,epsf,version2]{revtex}
\begin{document}
\draft

\title
\bf
Self-Organized Criticality and Punctuated Equilibria   
\endtitle

\author
{Per Bak$^{1}$ and Stefan Boettcher$^2$}
\instit
$^1$Department of Physics, Brookhaven National Laboratory,
Upton, NY 11973\\
$^2$Center for Theoretical Studies of Physical Systems,
Clark Atlanta University, Atlanta, GA 30314\\
\endinstit

\abstract 
Many natural phenomena evolve intermittently, with periods of tranquillity
interrupted by bursts of activity, rather than following a smooth
gradual path. Examples include earthquakes, volcanic eruptions, solar
flares, gamma-ray bursts, and biological evolution. Stephen Jay Gould and
Niles Eldredge have coined the term "punctuated equilibria" for this
behavior. We argue that punctuated equilibria reflects the tendency of dynamical systems
to evolve towards a critical state, and review recent work on simple models.
A good metaphoric picture is one where the systems are temporarily trapped
in valleys of deformable, interacting landscapes. Similarities with spin glasses are pointed out.
Punctuated equilibria are essential for the emergence of complex phenomena.
The periods of stasis allow the system to remember its past history; yet
the intermittent events permit further change.

\medskip
\noindent
KEYWORDS: Self-Organized Criticality, Punctuated Equilibrium,
Avalanches, Landscapes, Glassy Dynamics.
\endabstract

\noindent
e-mail: $^1$bak@cmth.phy.bnl.gov and $^2$stb@hubble.cau.edu
\medskip

\section{Introduction}
At the time of Darwin, the geophysicist Charles Lyell \cite{Lyell} formulated the
philosophy of uniformitarianism, according to which everything should
be explainable in terms of the currently observable processes working
at all times and all places with the same intensity. No new principles
need to be established for the great and the lengthy. Lyell took for
granted that this would lead to a smooth type of dynamics.

Darwin accepted Lyell's uniformitarian vision, and believed that his mechanism
for evolution, random mutation followed by selection and proliferation of the
fitter variants, would necessarily lead to a gradual evolution. He even brought
a copy of Lyell's work with him on the {\it beagle}. He
went as far as to deny the occurrences of mass extinction events. Within
the uniformian view, mass extinctions must necessarily be caused by external
cataclysmic events, and most investigators have  assumed this without further
ado, looking for meteors, volcanic eruptions and flooding events as the cause
of mass extinction.

Lyell's view may seem surprising in light of the fact that large events dominate
his own science, geophysics. Large earthquakes killing thousands of people
occur with merciless regularity, as do volcanic eruptions and catastrophic floodings.
So what is wrong with his theory? Basically, the statement of uniformitarianism is a statement
of linearity. Small impacts have small effects. The combined effect of many small
impacts is a response with a Gaussian distribution, with vanishing probability
of large events. Systems in equilibrium are linear, so the underlying picture is
one where nature is in balance. However, in Physics we are aware that many dynamical
systems show non-equilibrium, non-linear behavior. In particular, large dynamical
systems are known to evolve to critical states, where the response to small impacts
may be enormous, reflecting a divergent susceptibility.

Indeed, evidence has been mounting that evolution in fact proceeds 
in a stepwise fashion with long periods of apparent tranquillity in the 
development of most species, interrupted by rapid intermittent activity.
The failure to find ``missing links'' in
the paleontological data and the occurrence of fluctuations of all sizes in
the extinction rates for species over time \cite{raup} have put the gradualistic belief into
question. The fact that extinction events occur over a broad range of scales,
including the smallest, 
indicates that a common intrinsic mechanism is at work.

Gould and Eldredge \cite{G+E} have proposed punctuated equilibrium as
a means to reinterpret the data, where species experience lengthy periods of
stasis in between short periods of rapid adaptation that leave little trace of
a missing link. While plausible, this theory does not explain
the dynamical origin of this phenomenon in many different systems.
 In recent years,
much effort has been spent on finding a process that generically leads to
punctuated equilibrium behavior
(for reviews, see Refs.~\cite{uniform},\cite{BP}).

We shall discuss a simple model of such an evolution process.
A fitness value is assigned to each species in the system. The fitness reflects the
species' ability to survive in its environment, consisting of
other species and external factors. The
fitness values, together with the interrelation of species (e. g. in a 
food chain), represent a complicated fitness landscape in which species co-evolve,
ruled by Darwinian principles. Local updating rules for the fitness
of each species leads the system into a state of self-organized criticality 
(SOC) \cite{BTW} with co-evolutionary avalanches of all sizes in the species ecology, and 
punctuated equilibrium behavior in the evolution of each species. Adjustments in the
fitness of one species imply changes in the fitness landscape experienced by
interacting species. We show that a hierarchical structured landscape 
emerges which stores the memory of past activity \cite{BoPa}. In this landscape we
find that fitness values are ``ultrametrically'' related, borrowing a phrase
known from the study of spin glasses. Disturbances dissipate 
slowly (like power laws), and avalanches exhibit aging behavior.

\section{Landscapes and Evolution Modeling}
A powerful metaphoric picture of biological evolution was proposed by 
S. Wright (for a review, see Ref.~\cite{wright2}) who
long ago introduced the idea of species evolving in a rugged fitness 
landscape with random mutation 
and differential selection towards higher fitness. This contrasts with
generally accepted ideas of Fisher \cite{fisher} and others, whose 
view of  evolution can be translated into one where the individual species
climb an infinite hill with a constant slope at a constant rate.
Even today, many models consist of single
species evolving in more sophisticated but static environments.
In our view, none of these scenarios contain any hint as to the origin
of the intricate complexity, involving the interaction between myriads of species,
which characterizes real Life. Few theories consider evolution as a process with many degrees
of freedom, representing many different species.

One class of models that do include the many body
aspect of evolution has been proposed by Kauffman and Johnson \cite{K+J}.
These so-called NKC-models are remarkable because they introduce the
idea of a dynamically evolving fitness landscape. Here the fitness 
landscape is no longer merely a rigid stage for the individual species, but is 
itself formed by the ever-changing properties of other species, providing a 
model of co-evolution. Unfortunately, these models are
very complicated, and interesting features (such as criticality and
punctuated equilibrium) only emerge under very special circumstances, specifically
when a parameter characterizing the roughness of the landscape is tuned to
a critical point at a phase transition.

A much simplified model of evolution in which many species 
co-evolve in a dynamically formed fitness landscape has been proposed recently 
by Bak and Sneppen \cite{B+S}. Species are considered to be perfectly
distinguishable entities whose overall fitness with respect to their
environment can be simply characterized by a single number on the unit
interval. This value can be thought of as the fitness value of a local peak 
in a fitness landscape. 
Species are placed on a lattice, which could for instance be a one dimensional
food chain. Each species interacts with its neighbors. 
At each time step 
the ``weakest'' species, that is the one with the lowest fitness,
 is ``mutated'' (following Darwinian principles
\cite{Darwin}), or, equivalently, eliminated and replaced by another species. 
This process is effectuated by replacing the fitness value with a random number drawn from a flat
distribution between 0 and 1. 
This event in turn forces changes in the fitnesses of
interrelated species. Their fitness landscape has changed by no fault of their own!
In the model we simply replace the fitness values of
all nearest neighbors with new random numbers from the same
distribution. 

The most important consequence of this model is that no matter what the
initial state, the system evolves inevitably (self-organizes) into a 
``critical'' state in which correlations in space and time between 
events are distributed without any characteristic scale except for the 
system size itself (i. e. the number $N$ of
species in the system). The robustness of this macroscopic behavior with
respect to all but the most drastic modification in the local update rules is
necessary for any model that is not to rely on ``divine intervention''.  A
snapshot of a typical finite system near the critical state is shown in
Fig.~\ref{snapshot}. All but a few of the fitness values are distributed
evenly above a certain critical fitness value $\lambda_{\rm c}$, leaving a gap
below. In fact, a rigorous equation that describes the formation of the gap
from an initial condition consisting of evenly distributed numbers to this critical 
state has been derived (``gap-equation'') \cite{PMB} \cite{review}. The fitness values above this
threshold $\lambda_{\rm c}$ represent species that are very unlikely to ever
become the weakest species in the system. They have reach stasis, i. e. they
are in an apparent
equilibrium that will only be punctuated when a weak neighboring species
undermines their adaptation to the environment. The species with fitness values below this
threshold value are the most active, i. e, most likely to mutate or go extinct,
since they are the most likely to become
a global minimum at some time step. These active species undergo a rapid sequence 
of changes until they and their neighbors collectively reach high 
fitness values to regain stasis, where the network of species is in a state of temporary balance. 

The active species form avalanches of
all sizes which represent those fluctuations that are found in real
evolutionary activity. An avalanche here consists of all such active sites 
between two consecutive points in time when all values are above threshold.
If we count the returns of activity to any one species in the critical
state, we find a ``devil's staircase'', Fig.~\ref{devil}, with plateaus of
stasis (distributed according to a power law) punctuated by short periods of
rapid activity. Thus, this model provides a dynamical explanation of
punctuated equilibrium emerging from Darwinian principles. 

Many interesting features have been unearthed about this and other models of
SOC, most of which have been summarized in Ref.~\cite{review}. Here we want
to focus on a variety of aspects in the Bak-Sneppen model that are related to
the complicated, ever-changing fitness landscape which each species faces. 

%xxx Moved headline up one paragraph and changed title
\section{Fitness Landscapes, Avalanche Hierarchies and Glassy Behavior}
Each species in the system is frozen into a state of
stasis, except for a vanishingly small number of time steps (see the devil's
staircase in Fig.~\ref{devil}) when the local changes in the fitness landscape 
require adaptive activity. This change in the local fitness landscape is
caused by a similar adaptive move in a neighboring species at some previous
time, and so on. If a smaller fitness value has been assigned to a species, the 
barrier against motion in this landscape has been 
lowered.  But if a species and all its neighbors have fitness values anywhere 
above $\lambda_{\rm c}$, further spontaneous mutations will not take place
 until the activity returns to that species.
The system is attracted to a
state in which all fitness values are somewhere above threshold. There is a
continuum of such states. Different states are separated by intermittent activity in form of
avalanches that rearrange the system in a trial-and-error search.

The avalanches are hierarchically structured. Consider the time signal of
such an avalanche in form of the fitness value $\lambda_{\rm min}(s)$ of the 
weakest species as a function of update time $s$, Fig.~\ref{forw}. The distribution of
avalanches is given by a power law. Considering the infinite, critical 
avalanche, we can regard every update as a starting point of 
a sub-avalanche labeled by $\lambda_{\rm min}(s)$ which ends at the first time 
when $\lambda_{\rm min}(s')>\lambda_{\rm min}(s)$ for a $s'>s$. Clearly, a
$\lambda_{\rm min}(s)$ avalanche can only end when all of its sub-avalanches
with $\lambda_{\rm min}(s'')<\lambda_{\rm min}(s)$ for all $s<s''<s'$ have ended.
And a $\lambda_{\rm c}$ avalanche can only end when all of its sub-avalanches
have ended, and so on. Thus, a picture of hierarchically constrained dynamics emerges in
which faster degrees of freedom block slower ones, similar to the
phenomenological description of slow relaxation in glassy systems given by
Palmer, Stein, Abrahams, and Anderson in Ref.~\cite{PSAA}. This
was pointed out by Ref.~\cite{BoPa2}.
 Actually, The similarity with spin glasses is 
pretty straightforward. The fitnesses can be thought of as 
barriers against further action, i. e. the barriers that atoms have to traverse
in order to get to a better energy minimum. Once the atom jumps, the barriers of neighbor atoms are affected. The
duration of avalanches in the self-organized critical state is found to be
broadly distributed, following a power law, as a mark of this constrained
relaxation process.

By design, all active fitness values (in general, all values that had been
active at some time) are ultrametrically related \cite{ultram}. 
By choosing a site with a
specific (parental) fitness value at each update to create new (offspring) 
fitness values in its neighborhood, a causal link between previous and future 
fitness values is established, similar to relations in a family tree. In the
evolving avalanche, for instance, any two active fitness values are related
in a unique way to a closest common ancestor such that the ultrametric
relation between any three of them holds, see Fig.~\ref{tree}. It is found
that the ultrametric separation between two consecutively updated fitness
values is also distributed in a power law, another indication of
the highly correlated, slow relaxation process in evolving avalanches.

This ultrametric structure is reminiscent of earlier models for relaxation in
spin glasses due to Ogielski and Stein, and Schreckenberg \cite{OSS}. There the 
endnodes on a fixed ultrametric tree are the states between which a random
walker proceeds, by following jump probabilities that are chosen with respect
to the
ultrametric distances between states. The jumps of the walker correspond
to the jumps in consecutive activity in the Bak-Sneppen model. But in the
Bak-Sneppen model, the jump probabilities and a (random) ultrametric tree 
structure emerge {\it dynamically}. In fact, the jump probabilities in
this model are closely linked to the extremal process of always choosing
the smallest fitness value in each update, and would become short-range,
if an arbitrary active sites were to be updated instead.

Since the Bak-Sneppen model can be thought of as the dynamics in a rugged
landscape, it is
not surprising to find phenomena in the Bak-Sneppen model that are
usually associated with spin glasses 
\cite{S+N}. For instance, it can be shown (see below) that the updating rules 
lead to a build-up of memory of past activity over all scales,
resulting in equations similar to those for correlations in spin glasses \cite{C+K}.
Furthermore, detailed studies of the temporal activity reveal that
avalanches age \cite{Bouchaud}: the probability $P(t,t_w)$ for the activity to return
for the first time to a site at time $t+t_w$ 
depends on the age $t_w$ of the avalanche at the previous pass \cite{BoPa3}.
%xxx taken out figs
The connection between hierarchically
structured dynamics and aging has been made previously, again
in the context of spin glasses \cite{Sibani}.

\section{The Multi-Trait Model}
As a realistic extension of the Bak-Sneppen model, one may consider making the
survivability of each species conditional upon a number ($M$) of
independent traits associated with the different tasks that it has to
perform \cite{BoPa}.  We find that such a model preserves the generic 
properties (such as punctuated equilibrium) of the Bak-Sneppen model 
to which it reduces for $M=1$. 
In the limit $M\to\infty$, this model allows rare analytical
insights into the dynamics of avalanches in the SOC state. 

{}For $M=\infty$, 
an exact evolution equation for avalanches can be derived directly from 
the microscopic update rules. The SOC state emerges as a particular
point in the equation where simple diffusive behavior is replaced by 
long-range memory which the avalanche develops. Its solution 
provides a set of exact scaling coefficients that explicitly verifies
many of the proposed scaling relations for this class of extremal models
\cite{scaling}. Furthermore, it elucidates the subtle properties
that evolve from the irreversible dynamics, such as the Levy-flight 
distribution of adaptive activity and the ultrametric structure of 
the avalanche. These features are intimately connected to the distribution
of ``backward'' avalanches which is obtained in closed form. For a derivation
of these results, see Refs.~\cite{BoPa,BoPa2}.  

As in the Bak-Sneppen model, a species is represented by 
a single site on a lattice.  But in the multi-trade model the 
collection of traits for each
species is represented by a set of $M$ numbers in the unit interval. A
larger number represents a better ability to perform that particular 
task, while smaller numbers pose less of a barrier against mutation. 
Therefore, we ``mutate'' as before at every time step the smallest number 
in the entire system. Now, each neighboring species has {\it one} of its 
$M$ numbers replace. Which one of the $M$ numbers
is selected for such an update is determined at random, since we assume
that a mutation in the traits of one species can lead to an adaptive
change in {\it any one} of the traits of a neighboring species.  Thus, 
on a nearest-neighbor site, any number that is part of the avalanche has
a $1/M$ chance to be eliminated from the avalanche. In a nutshell,
the model is solvable for $M=\infty$ because fitness values below threshold
in the avalanche 
evolve {\it statistically independent} from one another and can be 
eliminated only by becoming the global minimum. But for {\it any} $M\leq\infty$,
the interaction between the fitnesses of species leads to a chain reaction
of coevolution that inevitably evolves to a self-organized critical state
as in the Bak-Sneppen model, $M=1$. It is important to note that
all the properties of the Bak-Sneppen model that indicate a rugged
fitness landscape are present also for arbitrary $M$ including
$M=\infty$. While this multi-trait model shares some features with earlier
mean-field versions \cite{meanfield}, it preserves nontrivial spatio-temporal 
correlations including the Devil's Staircase and punctuated equilibria
much like the Bak-Sneppen model.

To describe these spatio-temporal correlations, we define
$F(r,s)$ to be the probability for an avalanche in the SOC state
to survive precisely $s$ steps and to have affected a
particular site of distance $r$ from its origin. Due to the statistical
independence of active fitness values, one can find an exact evolution
equation for $F(r,s)$ at $M=\infty$.

In Ref.~\cite{BoPa} that equation was used to show that the system becomes
``critical'' with power laws for the avalanche duration and for
the spatial extent of avalanches. Thus, at least two scaling
coefficients can be found:
\begin{eqnarray}
F(r=0,s)\sim {s^{-{3\over 2}}\over\sqrt{\pi}}\quad(s\gg 1),\qquad{\rm and}
\qquad -\partial_r \sum_{s=0}^{\infty}F(r,s)\sim 24\, r^{-3} \quad (r\gg 1),
\end{eqnarray}
i. e. $\tau=3/2$ and $\tau_R=3$.
All other critical coefficients of the model can be determined from 
these two via scaling relations \cite{scaling}. For instance, 
the spatial extent of the activity in the SOC state spreads in a subdiffusive
manner, $r\sim s^{1/D}$, where $D=4$ is the avalanche dimension
given by $\tau_R -1 = D(\tau -1)$.
Below we verify the scaling relation for $D$ explicitly 
by calculating the diffusion behavior directly \cite{BoPa2}.
Other critical exponents that can be determined explicitly are 
$\gamma=1$, $\nu=\sigma=1/2$, $\tau_f^{all}=2$, and
$\tau_b^{all}=3/2$ (For a definition of these exponents see
Ref.~\cite{scaling}.

In a long-lived avalanche, each site is visited many times,
leading to punctuated equilibrium behavior, characterized by
the distribution of first returns
of the activity to a given site, $P_{\rm FIRST}(s)$.
It has been found that $P_{\rm FIRST}(s) \sim s^{-\tau_{\rm FIRST}}$ for 
large $s$ with $\tau_{\rm FIRST}= 2-d/D$. For $M\to\infty$ 
we find that  $\tau$ and $\tau_R$, and hence $D$, do not change with 
dimension $d$, and it is 
\begin{eqnarray}
\tau_{\rm FIRST}=2-d/4\quad(d\leq 4).
\end{eqnarray}
Thus, for $d=1$ we predict  $\tau_{\rm FIRST}=7/4$, and we find
numerically $\tau_{\rm FIRST}=1.73\pm 0.05$ \cite{BoPa}.

In Ref.~\cite{BoPa2} it is shown for sufficiently large $r$ and $s$ that 
\begin{equation} 
{\partial F(r,s)\over \partial s}\sim {1\over 2}\nabla_r^2 F(r,s) +
{1\over 2} \int_0^s V(s-s')F(r,s')ds' , 
\label{nonlocaleq} 
\end{equation} 
which is a ``Schr\"odinger'' equation in imaginary time for $F(r,s)$ with a
nonlocal kernel $V(s)=F(r=0,s)-2\delta(s)$.
This nonlocal potential with the integral kernel $V(s)$ contains all of
the history dependence of the process.  In its absence the system would
be purely diffusive with a Gaussian tail $F \sim e^{-r^2/s}$. In its
presence the probability to have reached a site at distance $r$ at time
$s$ gets contributions from avalanches that reached $r$ at earlier times
$s'<s$.  These contributions are weighted according to $V(s-s')$ which
has a power-law tail, representing the memory of the avalanche of previous
activity over all time scales.  The ultrametric tree structure of avalanches 
shows that they can be divided into sub-avalanches.
Avalanches contributing to $F(r,s)$
consist of sub-avalanches, one of which reaches $r$ in time $s'$ while
the other's combined duration is $s-s'$.  The sub-avalanche structure
gives a hierarchy of time scales. This changes the relaxation dynamics 
to be non-Gaussian. The form of Eq.~(\ref{nonlocaleq}) is reminiscent of
equations that describe slow dynamics and aging in spin glasses
\cite{C+K}.

Using a Laplace transform and steepest-decent
analysis of the inverse transform integral, one finds a form for the
propagator that might be rather general for systems with SOC \cite{PaBo}:
\begin{equation} 
F_{\lambda_{\rm c}}(r,s) \sim 
\exp\left[-C \left({r^D\over
s}\right)^{1\over D-1}\right] \qquad\left(r^D \gg s \gg 1\right).
\label{nongaussian} 
\end{equation} 
Assuming that for any system with SOC the history dependence is given by
$V(s)\sim s^{-\alpha}$, it is $D=2/(1-\alpha)$. For $M\to\infty$, 
it is $\alpha=3/2$, i. e. $D=4$, and the constant $C=3/4$.
Since $1<\alpha<2$, diffusion is slowed down ($D>2$) because the activity 
has a tendency to revisit sites, and the system remembers these previously 
visited sites. 

Thus, an intricate structure emerges in the self-organized critical state of
this intriguingly simple model.
At many places we eluded to similar phenomena observed in 
spin glasses. In fact, although their connection is not at all
clear, the Bak-Sneppen model might provide a more accessible setting to
study some of the phenomena shared with much more complicated spin glass
systems.

\figure{\label{snapshot}
Snapshot of the stationary state in the one dimensional
Bak-Sneppen model.
Except for the avalanche which consists of small fitness values in 
a localized region, almost all the fitness values in the system are 
above a self-organized threshold $\lambda_c$.}
\epsfxsize=350pt
\epsfysize=450pt
\epsffile{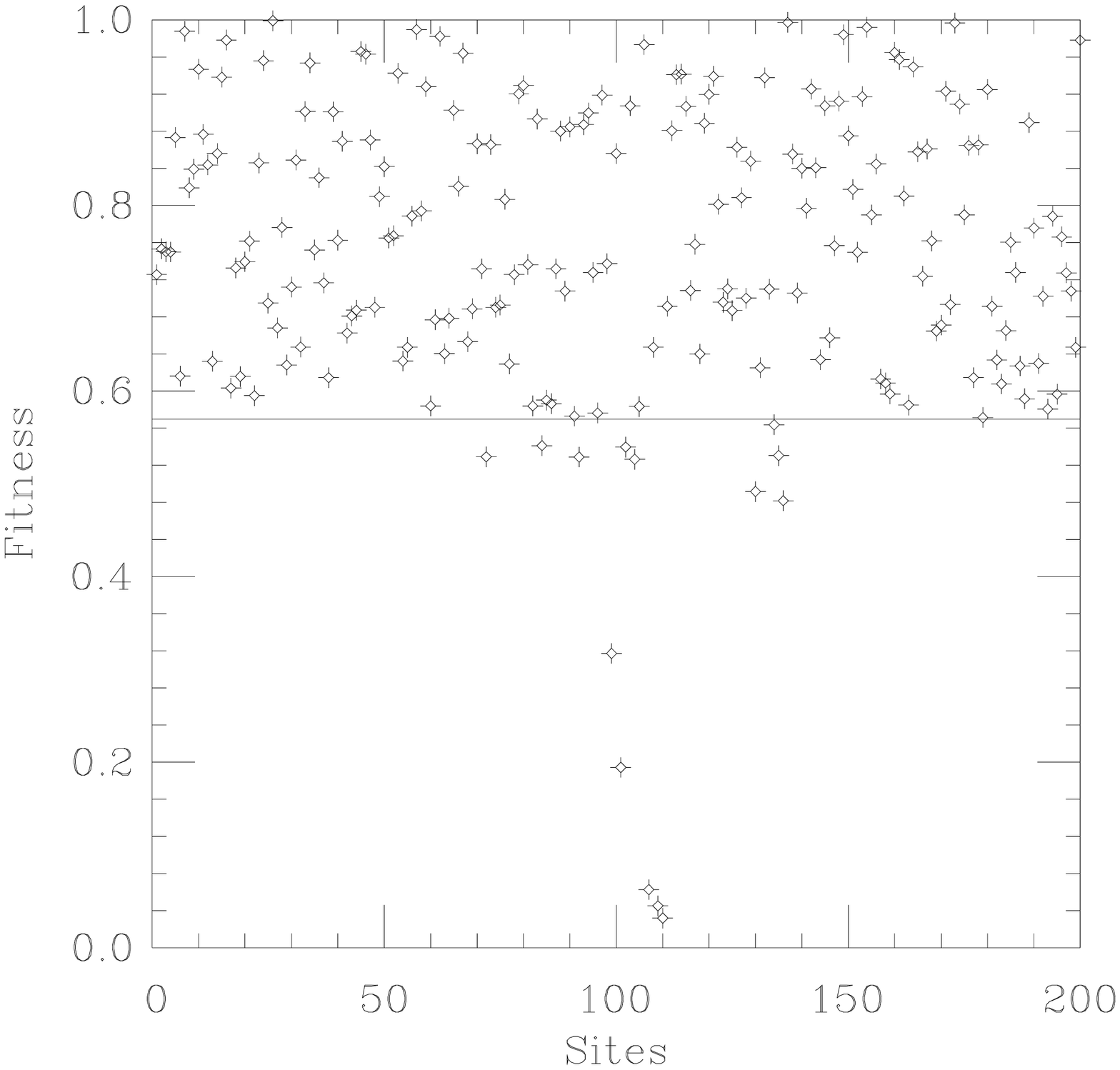}
\newpage

\figure{\label{devil}
Punctuated equilibrium behavior for the evolution of a single  species
in the one-dimensional $M=\infty$ model.  The vertical axis is the
total number of returns of the activity to some site as a function of
time $s$.  Note the presence of plateaus (periods of stasis)  of all sizes.
The distribution of plateau sizes scales as $s^{-7/4}$.}
\epsfxsize=300pt
\epsfysize=400pt
\epsffile{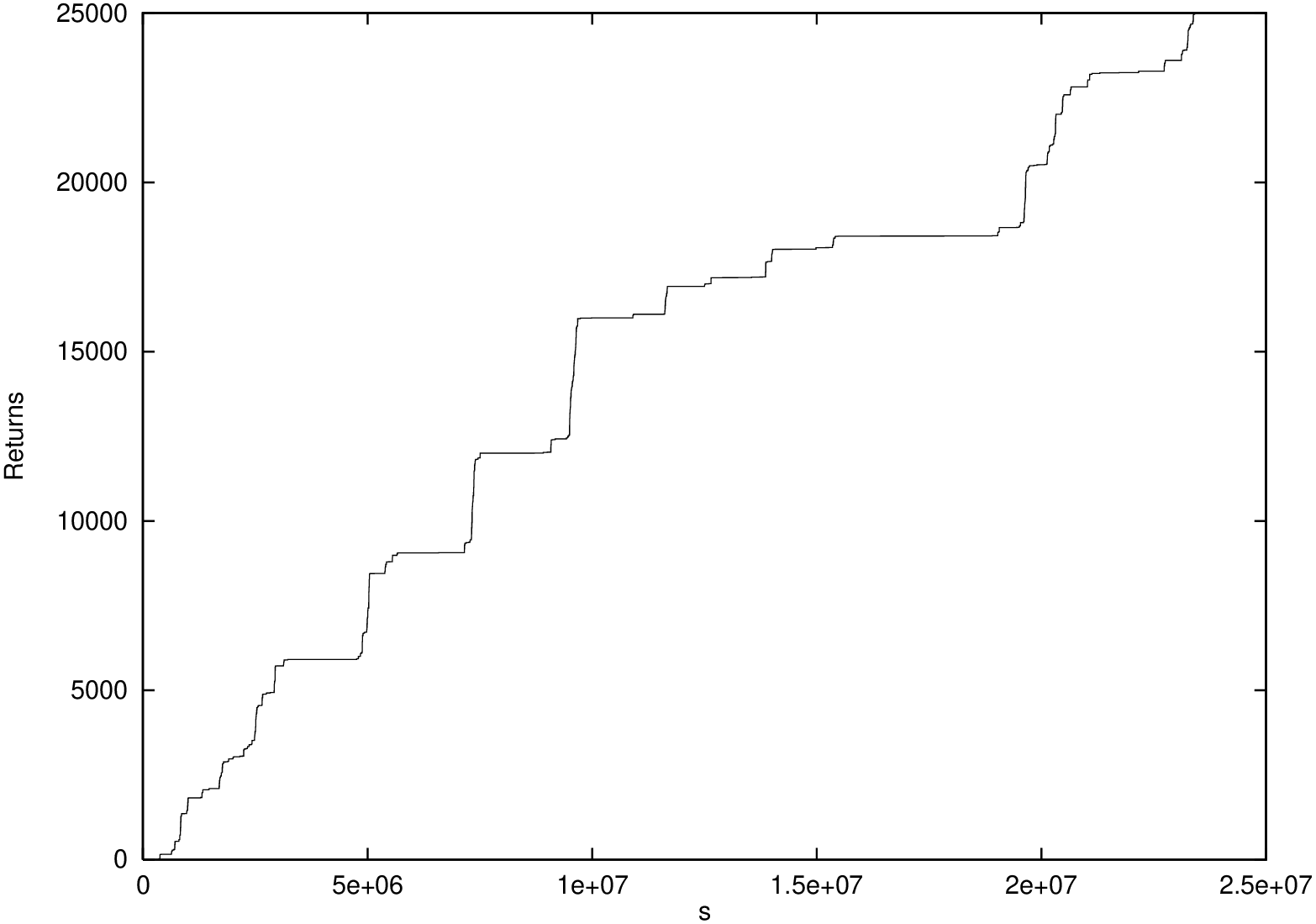}
\newpage

\figure{\label{forw}
Sequence of minimal random numbers 
$\lambda_{\rm min}(s)$ chosen for an update at time $s$ in a 
$\lambda_{\rm c}$ avalanche for $M=\infty$. The durations avalanches
within the hierarchy of 
$\lambda$ avalanches is indicated by forward arrows, where 
$\lambda=\lambda_{\rm min}(s)$. Longer avalanches with larger values of
$\lambda$ contain many shorter
avalanches which have to finish before the longer avalanche can terminate.
Note that an update punctuates any $\lambda$ avalanche with
$\lambda\leq \lambda_{\rm min}(s)$.}
\epsfxsize=350pt
\epsfysize=450pt
\epsffile{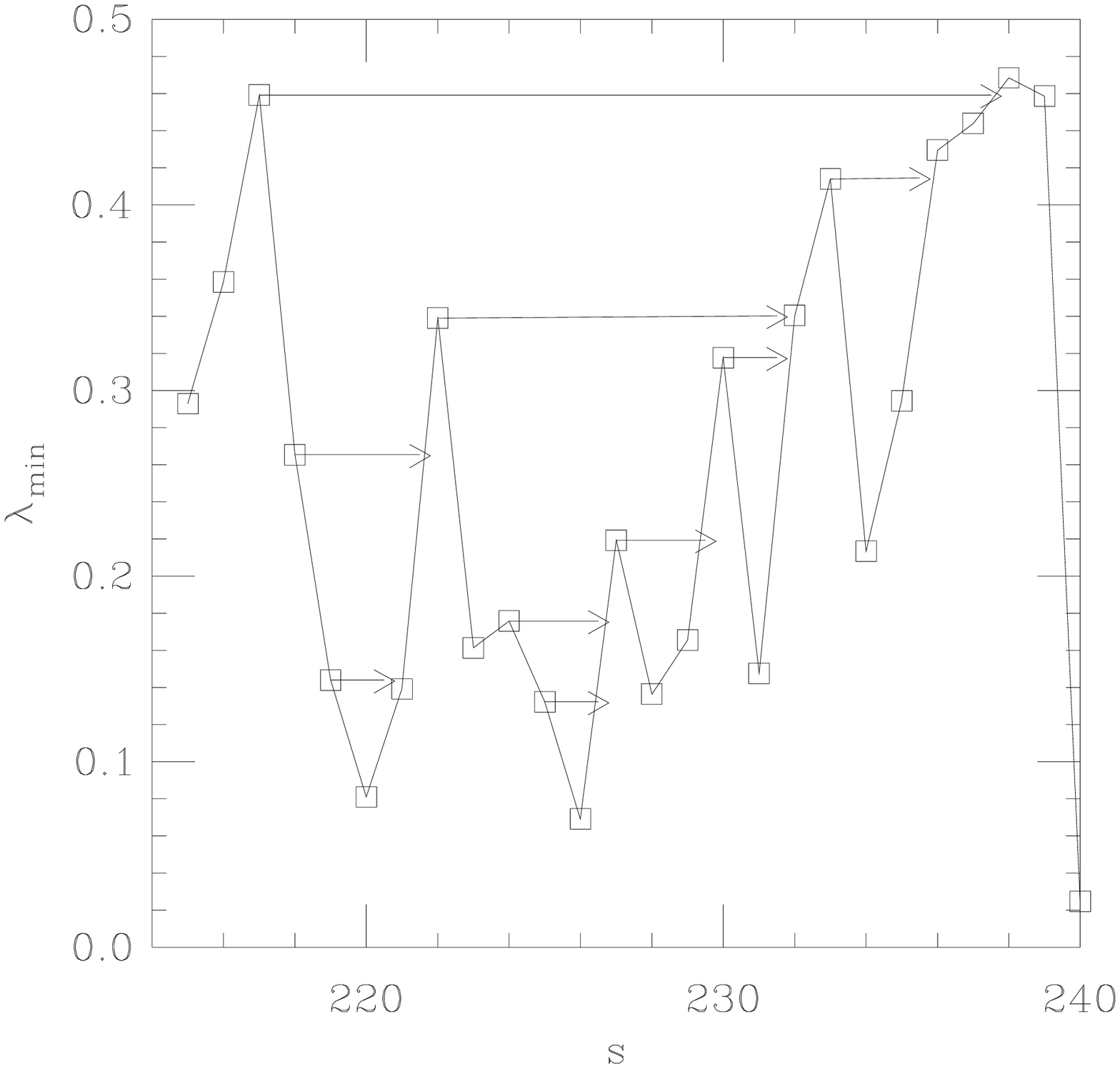}
\newpage

\figure{\label{tree}
Ultrametric tree structure.  At any given time, indicated by the vertical
axis, all of the active sites below threshold 
have an ancestry which forms a tree.
The ultrametric distance between any pair is the distance back in time
to the first common ancestor.}
\epsffile[0 600 100 800]{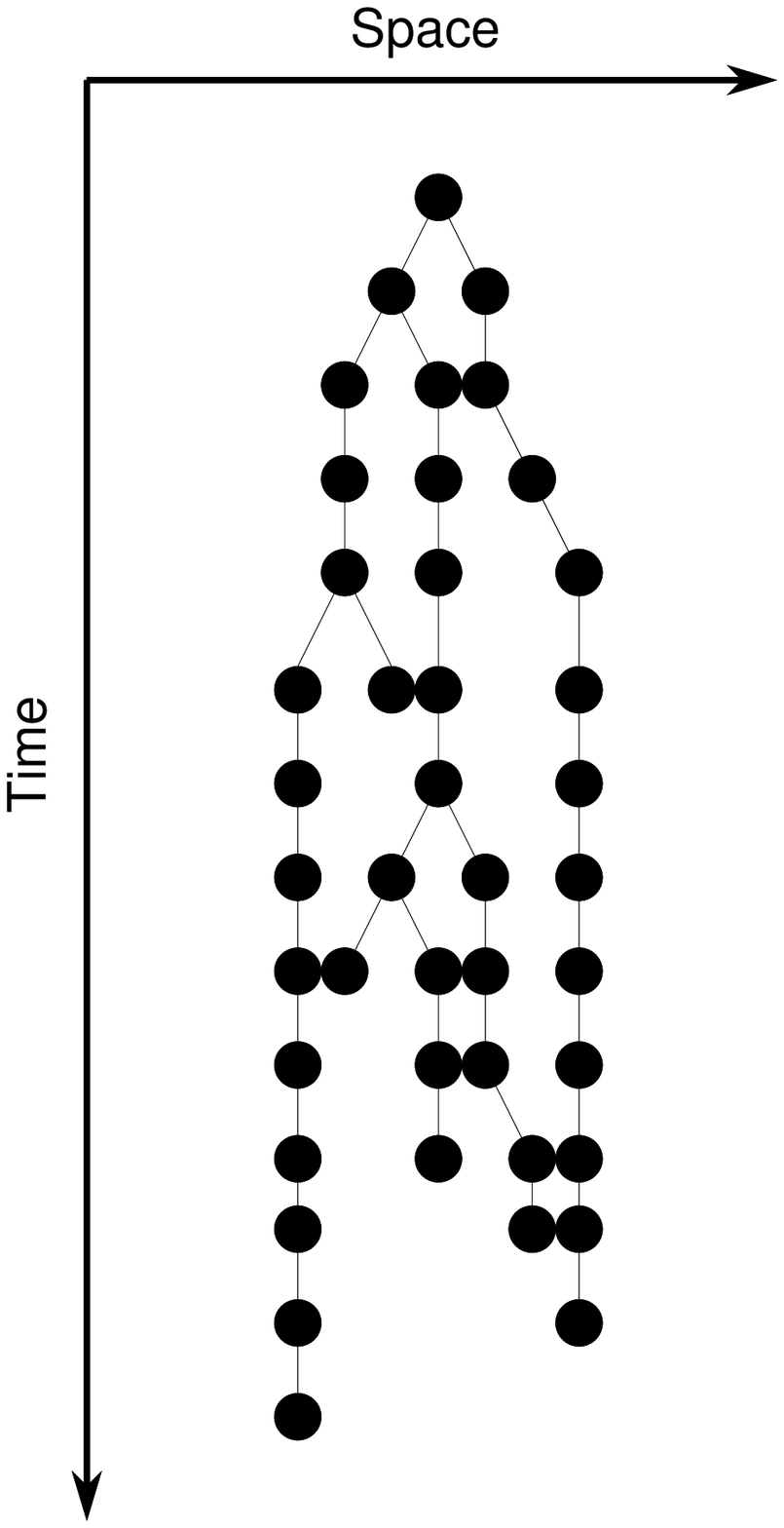}
\newpage

\end{document}